\documentclass[11pt,twoside]{article}
\usepackage{asp2010}

\def\rxj {RX~J1856.5$-$3754}

\resetcounters

\begin{document}

\title{A study of the long term variability of \rxj\ with XMM-Newton}
\author{S.~Mereghetti,$^1$ N.~Sartore,$^1$ A.~Tiengo,$^{2,1}$ A.~De~Luca,$^1$
R.~Turolla,$^3$ and F.~Haberl$^4$
\affil{$^1$INAF, IASF-Milano, via E. Bassini 15, 20133, Milano, Italy}
\affil{$^2$IUSS, viale Lungo Ticino Sforza 56, 27100, Pavia, Italy}
%
%
\affil{$^3$Padova University, via Marzolo 8, 35131, Padova, Italy}
\affil{$^4$MPI f{\"u}r extraterrestrische Physik, Giessenbachstra{\ss}e, 85748 Garching, Germany }
}

\begin{abstract}
We report on  a detailed spectral analysis of all the available XMM-Newton data of  \rxj , the
brightest and most extensively observed nearby, thermally emitting neutron star. Very small
variations ($\sim$1-2\%) in the single-blackbody temperature are detected, but are probably due to
an instrumental effect,  since they correlate with the position of the source on the detector.
Restricting the analysis to a homogeneous subset of observations, with the source  at the same
detector position, we place strong limits on possible spectral or flux variations from March 2005
to present-day. A slightly higher temperature (kT$\sim$61.5 eV, compared to the average value
kT$\sim$61 eV) was instead measured in April 2002. If this difference is not of instrumental
origin, it implies a rate of variation of about 0.15 eV yr$^{-1}$ between April 2002 and March
2005. The high-statistics spectrum from the selected observations is well fit by the sum of two
blackbody models, which extrapolate to an optical flux level in agreement with the observed value.
\end{abstract}

\section{Introduction}

The X-ray Dim Isolated Neutron Stars (XDINSs,  see \citet{tur09} for a review) are  a small group
of nearby, $d\lesssim300\,\rm pc$, thermally-emitting neutron stars, characterized by temperatures
$kT^\infty\sim50-100\,\rm eV$, luminosities $L_X\sim10^{31}-10^{32}\,\rm erg\,s^{-1}$, and spin
periods in the 3--12 s range. They are radio-quiet and have very faint optical/UV counterparts,
$m_V\sim26-27$.   Their spin-down rates ($\sim10^{-14}-10^{-13}\,\rm s\,s^{-1}$) imply magnetic
fields $B\sim10^{13}-10^{14}\, \rm G$, in good agreement with those inferred from the broad
spectral features observed in some of these sources.

Only one of the XDINSs, RX J0720.4$-$3125, showed significant long term spectral and flux
variations \citep{dev04,hoh12a}, for which several interpretations were proposed, including
precession of the neutron star and changes induced by a glitch \citep{hab06,van07}. No evidence of
variability has been reported in the other six members of this class, but most of them have not
been observed very frequently and have lower fluxes than RX J0720.4$-$3125, therefore the derived
limits are not very constraining.

On the other hand, \rxj\ is the brightest XDINS and it has been  routinely observed, almost twice
per year since 2002   for calibration purposes, by the XMM-Newton X-ray satellite. The resulting
large amount of data allows us to investigate its spectral evolution on time scales from months to
$\sim$10 years, but a reliable interpretation of these data requires to carefully consider the
stability of the detectors and any calibration issue that might affect the results.

\begin{figure*}[bht!]
\begin{center}
\includegraphics[width=0.75\textwidth,angle=0]{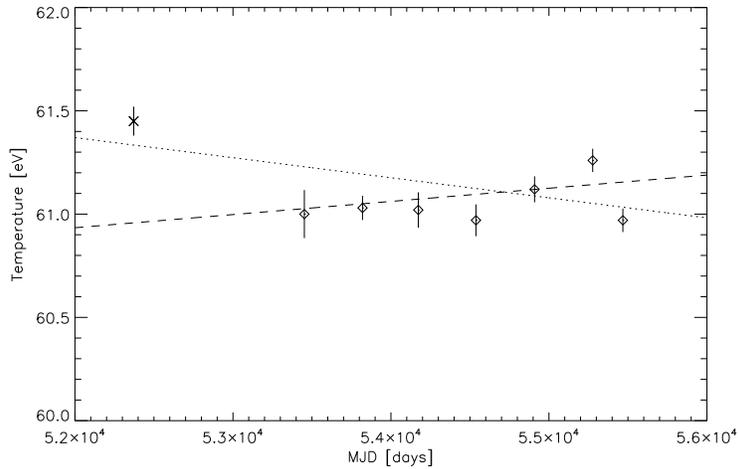}
\caption{Long term evolution of the BB temperature from a set of homogeneous observations with the
source at the same position on the EPIC pn detector. The dotted line is a linear fit to all the
points. The dashed line represents a linear fit excluding the first point (April 2002).}
\label{fig-kt}
\end{center}
\end{figure*}

\begin{figure*}[bht!]
\begin{center}
\includegraphics[width=0.7\textwidth,angle=-90]{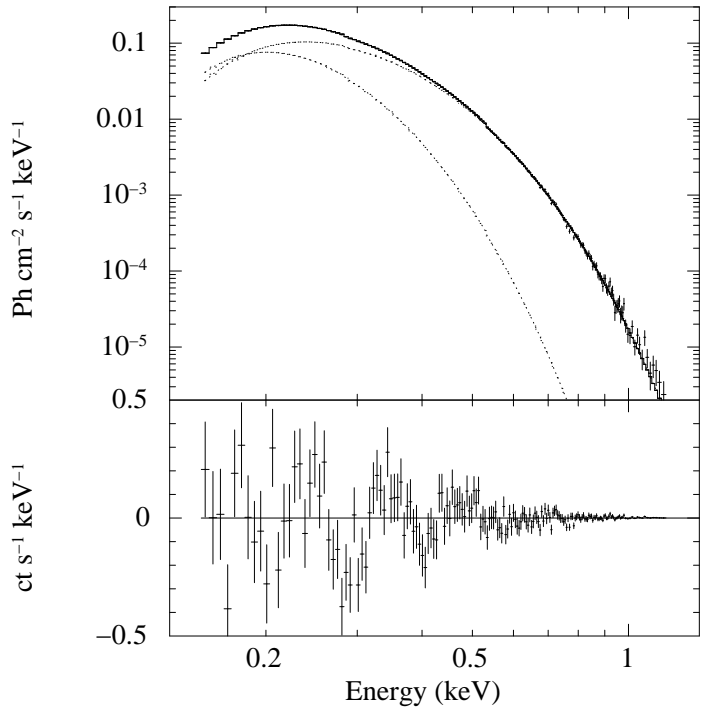}
\caption{Spectrum of \rxj\ extracted from a homogenous set of observations yielding a total
exposure time of $\sim254$ ks.   The upper panel shows the best  fit with two blackbodies. The
features in the residuals (lower panel) have a width smaller than the pn energy resolution and are
most likely due to non perfect instrumental calibrations.} \label{fig-sp}
\end{center}
\end{figure*}

\section{Results}

We studied the spectral evolution of \rxj\ using all the available data obtained with the pn camera
of the EPIC instrument on board XMM-Newton. These consist of 18 observations carried out between
April 2002 and April 2012,  for a total exposure time  of more than 700 ks (after screening to
remove periods of high background). Complex physical models  have been used in the past to study
the X-ray emission from \rxj\ (see, e.g., \cite{tur04,ho07}), but, considering the lack of
prominent spectral features\footnote{in this respect \rxj\ is unique among the XDINSs}, and the
fact that we are mostly interested in the \textit{relative} spectral variations, we restricted our
analysis to fits with one or two absorbed blackbody (BB) models. All the fits were performed in the
0.15-1.2 keV energy range.

The fits with a single BB resulted in temperatures within $\sim$2\% of the average value
kT$\sim$61.5 eV. Although the very small statistical errors on the derived temperatures would imply
significant variability, a closer investigation indicated that the apparent variations are most
likely the results of instrumental effects (see \citet{sar12} for details). Due to the varying
satellite observing constraints, the source was observed at slightly different positions of the
EPIC pn detector. We found that the derived temperature correlates with the source position in
instrumental coordinates, in particular with the CCD column. This indicates that the effect is
probably caused by a non-corrected spatial dependence of the channel-to-energy conversion. Note
that this is a very small effect, which is relevant in this particular case of a high-statistics
spectrum of a very soft source, and it  does not affect most EPIC/XMM-Newton observations.

To minimize this instrumental effect, we restricted all the further analysis to a subset of eight
observations in which \rxj\ was located at approximately the same detector coordinates. In this way
we could set strong constraints on the relative variations in spectral shape and flux, but it must
be remembered that a systematic uncertainty of $\sim$1-2\% on the absolute values cannot be
eliminated.

The single BB model temperature is compatible with a constant during the last $\sim$5 years. A
linear fit gives a rate of temperature variation of $\sim0.023\pm 0.015\,\rm eV\,yr^{-1}$,
consistent with a constant at the $2\sigma$ level (see Fig.\ref{fig-kt}). A higher temperature was
measured in April 2002. If not caused by subtle alterations of the instrument response, this
difference would imply that also \rxj\ undergoes spectral changes, albeit much smaller than that
observed in RX J0720.4$-$3125, which underwent substantial and continuous changes in its spectral
properties during the years. The   temperature variation of \rxj,     $\sim$0.5 eV in three
years,\footnote{or less, considering the lack of coverage between April 2002 and March 2005}
corresponds to a rate of $\sim-0.15\,\rm eV\,yr^{-1}$.

The high-statistics spectrum ($\sim2\times10^6$ counts) obtained by summing the data of the
homogenous observations is shown in Fig. \ref{fig-sp}, with the two BB best fit model (a single BB
gives a significantly worse fit). The hard BB has a temperature of
$kT_h^\infty=62.4_{-0.4}^{+0.6}\,\rm eV$ with emission radius of
$R_{h}^\infty=4.7_{-0.3}^{+0.2}\,(d/120\,\rm pc)\,km$, while the soft BB has
$kT_s^\infty=38.9_{-2.9}^{+4.9}\, \rm eV$ and  $R_{s}^\infty=11.8_{-0.4}^{+5.0}\,(d/120\,\rm
pc)\,km$. The column density is   $N_{\rm H}=(12.9\pm2.2)\times 10^{19}\,\rm cm^{-2}$. We found no
convincing evidence for broad or narrow absorption features (see \citet{sar12} for details).

The contribution of the soft BB at optical wavelengths is four times larger than that of the hard
BB. This implies an overall increase of a factor $\sim$5 with respect to the optical flux expected
from the Rayleigh-Jeans tail of the hard BB alone. The resulting value is consistent, within the
uncertainties, with the observed optical/UV flux \citep{kap11}.

\section{Conclusions}

We have carried out a detailed analysis to investigate the long term spectral variability of the
brightest XDINS, reaching the limit set by the current uncertainties in the instrumental
calibration, and without finding evidence for relative variations larger than a few percent during
the last decade. We stress that the absolute values of the spectral parameters derived with
XMM-Newton for \rxj\ (and also for other very soft X-ray sources) must be taken with caution since
they can be affected by systematics errors at the level of few percent.

\acknowledgements We acknowledge the support of ASI/INAF (I/009/10/0).

\bibliography{axpsgr6}

\end{document}